\documentstyle[12pt,twoside]{article}
\def\greaterthansquiggle{\raise.3ex\hbox{$>$\kern-.75em\lower1ex\hbox{$\sim$}}}

\def\lessthansquiggle{\raise.3ex\hbox{$<$\kern-.75em\lower1ex\hbox{$\sim$}}}

\newcommand{\beq}{\begin{equation}}
\newcommand{\eeq}{\end{equation}}
\newcommand{\beqa}{\begin{eqnarray}}
\newcommand{\eeqa}{\end{eqnarray}}
\newcommand{\beqan}{\begin{eqnarray*}}
\newcommand{\eeqan}{\end{eqnarray*}}
\newcommand{\ba}{\begin{array}}
\newcommand{\ea}{\end{array}}

\newcommand{\ra}{\rightarrow}

\newcommand{\wt}{\widetilde}

\newcommand{\A}{{\cal A}}

\newcommand{\F}{{\cal F}}
\newcommand{\G}{{\cal G}}

\newcommand{\M}{{\cal M}}

\newcommand{\U}{{\cal U}}

\newcommand{\st}{\stackrel}

\def\nz{\ifmmode {I\hskip -3pt N} \else {\hbox {$I\hskip -3pt N$}}\fi}
\def\zz{\ifmmode {Z\hskip -4.8pt Z} \else
       {\hbox {$Z\hskip -4.8pt Z$}}\fi}
\def\qz{\ifmmode {Q\hskip -5.0pt\vrule height6.0pt depth 0pt
       \hskip 6pt} \else {\hbox
       {$Q\hskip -5.0pt\vrule height6.0pt depth 0pt\hskip 6pt$}}\fi}
\def\rz{\ifmmode {I\hskip -3pt R} \else {\hbox {$I\hskip -3pt R$}}\fi}
\def\cz{\ifmmode {C\hskip -4.8pt\vrule height5.8pt\hskip 6.3pt} \else
       {\hbox {$C\hskip -4.8pt\vrule height5.8pt\hskip 6.3pt$}}\fi}
\def\au{{\setbox0=\hbox{\lower1.36775ex%
\hbox{''}\kern-.05em}\dp0=.36775ex\hskip0pt\box0}}
\def\ao{{}\kern-.10em\hbox{``}}
\def\lint{\int\limits}
\renewcommand{\baselinestretch}{1.5} 
\normalsize
\begin{document}
\bibliographystyle{plain}

\begin{titlepage}
\begin{flushright}
UWThPh-1998-65 \\
hep-th/9901165 \\
extended version
\end{flushright}
\vspace{2cm}
\begin{center}
{\Large \bf Global Path Integral Quantization 
\\[5pt]
of Yang-Mills Theory} \\[40pt]
Helmuth H\"uffel* and Gerald Kelnhofer**\\
Institut f\"ur Theoretische Physik \\
Universit\"at Wien \\
Boltzmanngasse 5, A-1090 Vienna, Austria
\vfill

{\bf Abstract}
\end{center}
\renewcommand{\baselinestretch}{1.0} 
\small
Based on a generalization of the stochastic quantization scheme 
recently
 a modified Faddeev-Popov path integral density for the 
quantization of Yang-Mills 
theory was derived, the modification consisting in the presence of 
specific 
$finite$ contributions of the pure gauge degrees of freedom. Due to the 
Gribov 
problem the gauge fixing can be defined only locally and the whole space 
of 
gauge potentials has to be partitioned into patches. We propose a global 
path 
integral density for 
the Yang-Mills theory by summing over all patches, which can be proven
to 
be
manifestly independent of the specific local choices of patches and
gauge 
fixing 
conditions, respectively. 
In addition to the formulation on the whole 
space of gauge potentials we discuss the corresponding global path
integral on 
the gauge orbit space relating it to the 
original Parisi-Wu stochastic quantization scheme and 
to a proposal of Stora, respectively.

\vfill
\begin{enumerate}
\item[*)] Email: helmuth.hueffel@univie.ac.at
\item[**)] Supported by "Fonds zur F\"orderung der wissenschaftlichen 
Forschung in \"Oster\-reich",  project P10509-NAW
\end{enumerate}
\end{titlepage}
\renewcommand{\baselinestretch}{1.5} 
\normalsize
The Faddeev-Popov \cite{Fadd} path integral procedure constitutes one of 
the most popular quantization methods for Yang-Mills theory and is
widely 
used in elementary particle physics. It is, however,  well known that at
a 
non perturbative level due to the Gribov ambiguity \cite{Gribov}  a
unique 
gauge fixing in the full space of gauge fields is not possible so that
the 
Faddeev-Popov path integral procedure is  defined only locally in field 
space.

Several attempts  were presented to generalize   the above approach 
in order to establish global integral 
representations. We especially point out the construction of a
regularized 
Feynman--Kac functional integral by \cite{asorey}, the use of
equivariant  
cohomological techniques by \cite{Stora}  and the method of 
implementing 
BRST invariance globally by \cite{Becchi}. 

It is our aim to present a quite different 
argumentation based on a recently introduced
generalization \cite{annals,annalsym}  of the 
stochastic quantization scheme \cite{Parisi+Wu,Damgaard+Huffel,Namiki}.

Let $P(M,G)$ be a principal fiber 
bundle with   compact structure group $G$ over
the compact Euclidean space time $M$. Let $\A$ denote the space of all
irreducible connections  on $P$ and let $\G$ denote the gauge group, 
which is given by all vertical automorphisms on $P$ reduced by the
centre
of $G$. Then $\G$ acts freely on $\A$ and defines a principal 
$\G$-fibration
$\A \st{\pi}{\longrightarrow} \A/\G =: \M$ over the paracompact 
\cite{Mitter} space
$\M$ of all inequivalent gauge potentials
with projection $\pi$. 
Due to the Gribov ambiguity the principal 
$\G$-bundle $\A \ra \M$
is not globally trivializable. 
 
From \cite{Mitter} it follows that there exists a locally finite open 
cover $\U
=\lbrace U_{\alpha} \rbrace$  of $\M$ together with a set of background 
gauge fields $\lbrace A_{0}^{(\alpha) } \in \A
 \rbrace$  such that 
\beq
\Gamma_{\alpha} = \{ B \in 
\pi^{-1} (U_{\alpha})|D^{*}_{A_{0}^{(\alpha)}} (B - 
A_{0}^{(\alpha)}) = 0\}
\eeq
defines a family of local sections of $\A \ra \M$. Here 
$D_{A_{0}^{(\alpha)}}^*$ 
is the
adjoint operator of the covariant derivative $D_{A_{0}^{(\alpha)}}$ with 
respect
to $A_{0}^{(\alpha)}$. 
Instead of analyzing Yang-Mills theory in the original field 
space $\A$  we 
consider the family of trivial  principal $\G$-bundles 
$\Gamma_{\alpha} \times \G \ra \Gamma_{\alpha}$, 
which are locally isomorphic to the bundle 
$\A \ra \M$, where the isomorphisms are provided by the maps
\beq
\chi_{\alpha} : \Gamma _{\alpha} \times \G \ra 
\pi^{-1}(U_{\alpha}), \qquad
\chi_{\alpha} (B,g) := B^g
\eeq
with $B \in \Gamma_{\alpha}$, $g \in \G$ and 
$B^g$ denoting the  gauge 
transformation of $B$ by $g$. 

Using this mathematical setting   we
start with the Parisi--Wu approach for the stochastic quantization
of the Yang--Mills theory in terms of the Langevin equation
\beq
dA = - \frac{\delta S}{\delta A} ds + dW.
\eeq
Here $S$ denotes the Yang--Mills action without gauge symmetry breaking
terms and without accompanying ghost field terms,
$s$ denotes the extra time coordinate 
with respect to which the stochastic process is evolving, $dW$ is the
increment of a Wiener process. 

Making use of the Ito stochastic calculus 
we locally transform  the Langevin equation (3) into the adapted
coordinates
$\Psi = \left( \ba{c} B \\ g \ea \right)$, perform special    
geometrically distinguished 
modifications \cite{annalsym} of its drift and diffusion term -thereby 
leaving expectation values of gauge invariant observables unchanged-
 and finally arrive at
\beq
d\Psi = \left[- \wt G_{\alpha}^{-1} \frac{\delta S_{\alpha}^{\rm tot}
}{\delta \Psi}
+ \frac{1}{\sqrt{\det G_{\alpha}}} 
\frac{\delta(\wt G_{\alpha}^{-1} \sqrt{\det G_{\alpha}
})}{\delta \Psi} \right] ds
+ \wt E_{\alpha} dW.
\eeq
Here $\wt E_{\alpha}$ and  
$\wt G_{\alpha}^{-1}  = \wt E_{\alpha} \wt E_{\alpha}^*$ 
denote a specific vielbein and a  (inverse) 
metric, respectively, which are associated  to  
the above mentioned modifications. $S_{\alpha}^{\rm tot}$
denotes a total Yang-Mills action  
\beq
S_{\alpha}^{\rm tot} = \chi_{\alpha}^* S + pr_{\G}^* S_{\G}
\eeq
defined by the original Yang-Mills action $S$
 and by $S_{\G} \in C^{\infty}(\G)$ 
which is an 
arbitrary functional on $\G$ such that $e^{-S_{\G}}$ is integrable 
with respect to  the invariant volume density
 $\nu = \sqrt{\det (R_g^* R_g)}$  
on $\G$. $R_g$  is the
differential of right multiplication
transporting any tangent vector in $T_g \G$ back to the identity 
$id_{\G}$  on $\G$,  
$pr_{\G}$ is the projector $\Gamma_{\alpha} \times \G \ra \G$.
 We recall furthermore that $\A$ admits a natural metric 
which gives rise to an induced metric $G_{\alpha}$ on 
$\Gamma_{\alpha} \times \G$ 
where
\beq
\det G_{\alpha} = \nu^2 \, (\det \F_{\alpha})^2 \,
(\det \Delta_{A_{0}^{(\alpha)}})^{-1}.
\eeq
$\F_{\alpha} = D_{A_{0}^{(\alpha)}}^* D_B$
denotes the Faddeev--Popov operator and $\Delta_{A_{0}^{(\alpha)}}^{-1}$
is the inverse of the covariant Laplacian 
$\Delta_{A_{0}^{(\alpha)}} = D_{A_{0}^{(\alpha)}}^*
D_{A_{0}^{(\alpha)}}$.

The Fokker--Planck equation associated to (4) can easily be deduced and
its (non-normalized) equilibrium  distribution is 
obtained  as \cite{annalsym} 
\beq
\mu_{\alpha} \, e^{-S_{\alpha}^{\rm tot}}, \quad 
\mu_{\alpha} = \sqrt{\det G_{\alpha}}.
\eeq
It is the basic idea of the  stochastic quantization scheme to interpret
an equilibrium limit of 
a Fokker--Planck distribution as  Euclidean path integral measure.
Although our result implies 
unconventional 
$finite$ contributions along the gauge group (arising from the 
$pr_{\G}^* S_{\G}$ term) it is equivalent to the usual 
Faddeev--Popov prescription for Yang--Mills theory. This follows from 
the fact that for expectation 
values of gauge invariant observables these contributions along the
gauge 
group are exactly canceled out due to the normalization of the path
integral, see 
below.
 We stress once more that 
due to the Gribov ambiguity the usual Faddeev--Popov 
approach as well as -presently- our modified version 
are valid only locally in field space. 

In order to compare  expectation values on different patches we 
consider the  diffeomorphism in the overlap of two patches 
\beq
\phi_{\alpha_1 \alpha_2} : (\Gamma_{\alpha_1} \cap
\pi^{-1}(U_{\alpha_2})) \times \G 
\ra 
(\Gamma_{\alpha_2} \cap \pi^{-1}(U_{\alpha_1})) \times \G
 \qquad
\phi_{\alpha_1 \alpha_2} (B,g) := 
(B^{\omega_{\alpha_2}(B)^{-1}}, g).
\eeq
Here $\omega_{\alpha_2} : \pi ^{-1}(U_{\alpha_2}) \ra \G$ is 
uniquely defined (see \cite{annalsym})  
by $A^{\omega_{\alpha_2}(A)^{-1}} \in \Gamma_{\alpha_2}$. 
To the  density  $\mu_{\alpha}$ there is associated  a corresponding 
twisted top  form on $\Gamma_{\alpha} \times \G$ 
(see e.g. \cite{Bott}) which for simplicity we denote by the same
symbol.
Using for convenience a matrix 
representation of $G_{\alpha}$ \cite{annalsym} we 
straightforwardly verify that
\beq
\phi_{\alpha_1 \alpha_2}^* \, \mu_{\alpha_2} = \mu_{\alpha_1} \, .
\eeq
This immediately implies  that in overlap regions the 
 expectation values of gauge invariant observables $f \in
C^{\infty}(\A)$ 
 are equal 
when evaluated in different  patches
\beq
\lint_{(\Gamma_{\alpha_2} \cap \pi^{-1}(U_{\alpha_1})) \times \G}  
{\mu_{\alpha_2} \, e^{-S_{\alpha_2}^{\rm tot}} \chi_{\alpha_2}^* f}  =
\lint_{(\Gamma_{\alpha_1} \cap \pi^{-1}(U_{\alpha_2})) \times \G}
{\mu_{\alpha_1} \, e^{-S_{\alpha_1}^{\rm tot}} \chi_{\alpha_1}^* f} .
\eeq

Suppose now that we consider a different locally finite cover 
$\lbrace U^{\prime}_{\beta} \rbrace$  of $\M$ 
together with a new set of background 
gauge fields $\lbrace A_{0}^{\prime (\beta)}
 \rbrace$ so that  
 a new family $\lbrace \Gamma^{\prime}_{\beta} \rbrace$ 
 of local sections,  as well as  new maps $\chi_{\beta}^{\prime}$, 
densities $\mu^{\prime}_{\beta}$, total actions 
 $S_{\beta}^{\prime \, \rm tot}$ and another  partition of unity
 $\gamma^{\prime}_{\beta}$ are given.
Applying the above integration formula 
in overlap regions we can prove furthermore that
 \beq
\lint_{\Gamma_{\alpha}  \times \G}  
{\mu_{\alpha} \, e^{-S_{\alpha}^{\rm tot}}
 \chi_{\alpha}^* (f \pi ^* (\gamma_{\alpha} \gamma^{\prime}_{\beta 
}))}  =
\lint_{\Gamma^{\prime}_{\beta}  \times \G}  
{\mu^{\prime}_{\beta} \, e^{-S_{\beta}^{\prime \, \rm tot}}
\chi_{\beta}^{\prime *} (f \pi ^* (\gamma_{\alpha}
\gamma^{\prime}_{\beta}))}.
\eeq
Finally we propose the definition of the global expectation value of a 
gauge 
invariant observable $f \in C^{\infty}(\A)$ by summing over all the
partitions 
$\gamma_{\alpha}$ such that 
\beq
\langle f \rangle = 
\frac{\sum_{\alpha} \lint_{\Gamma_{\alpha} 
\times \G} \mu_{\alpha} \, e^{-S_{\alpha}^{\rm tot}}
\chi_{\alpha}^* (f \pi ^* \gamma_{\alpha})}
{\sum_{\alpha} \lint_{\Gamma_{\alpha} \times \G} 
\mu_{\alpha} \, e^{-S_{\alpha}^{\rm tot}}
\chi_{\alpha}^*  \pi ^* \gamma_{\alpha} }
\eeq
(for a preliminary presentation of this result see 
\cite{Vancouver}). 
Due to (9) it is trivial to prove that the global 
expectation value $\langle f 
\rangle$ is independent of  the specific 
choice of the locally finite cover $\lbrace U_{\alpha} \rbrace$, 
of the  choice of the background 
gauge fields $\lbrace A_{0}^{(\alpha)} \rbrace$ 
and of the choice of the partition of unity $\gamma_{\alpha}$, 
respectively.

As already indicated in \cite{annalsym} these structures can equally be 
translated into the original field space $\A$. With the help 
of the partition of unity the locally 
defined densities $\mu_{\alpha}$ as well as $e^{-S_{\alpha}^{\rm tot}}$
can be pieced together to give a globally well defined 
twisted top form $\Omega$ on $\A$
\beq
\Omega := \sum_{\alpha}  \chi_{\alpha}^{-1 \, *}
(\mu_{\alpha} \,  e^{-S_{\alpha}^{\rm tot}}) \pi ^* \gamma_{\alpha}. 
\eeq
The  global expectation value (12) then reads
\beq
\langle f \rangle = 
\frac{ \int_{\A} \Omega \, f} 
{\int_{\A} \Omega }
\eeq
which due to the discussion from above is independent of all
the particular local choices.

In addition   to the global expressions (12) and (14)
 for the path integral in the whole 
 space of connections the stochastic quantization scheme also offers the 
possibility of deriving the corresponding  formulation on the gauge 
orbit space $\M$: We consider  the projections of either 
the original Parisi--Wu Langevin equation (3)  or of  the modified 
equation (4) onto 
the gauge invariant subspaces $\Gamma_{\alpha}$ described by the 
coordinate $B$; in both cases we obtain (see  \cite{annalsym})
\beq
dB = \left[ -(G_{\alpha}^{-1})^{\Gamma_{\alpha} \Gamma_{\alpha}}
\frac{\delta S}
{\delta B} + 
\frac{1}{\sqrt{\det G_{\alpha}}}
\frac{\delta((G_{\alpha}^{-1})^{\Gamma_{\alpha} 
\Gamma_{\alpha}}\sqrt{\det G_{\alpha}})}
{\delta B} \right]ds + E_{\alpha}^{\Gamma_{\alpha}} dW.
\eeq
Notice that in local coordinates $(G_{\alpha}^{-1})^{\Gamma_{\alpha} 
\Gamma_{\alpha}}$ is
the pullback of the restriction on $U_{\alpha}$ of the inverse of a
globally
defined metric on the gauge orbit space $\M$  induced by the natural
metric on $\A$. 
Since
the locally defined equations (15) are transforming covariantly under
the local
diffeomorphisms issued by the coordinate transformations, using
\cite{belo} it is
straightforward to check that their further projections onto $\M$
are yielding a globally defined stochastic process.

In direct analogy to our  
derivation of the local Fokker--Planck densities (7) we obtain   that
the 
 Fokker--Planck equation associated to the projected 
Langevin equations (15) 
  has an equilibrium distribution  given 
 by just the gauge invariant part of the densities  (7)
\beq
\det \F_{\alpha} \,
(\det \Delta_{A_{0}^{(\alpha)}})^{-1/2} \,
e^{-\chi_{\alpha}^* S}.
\eeq
By using (9) we can prove explicitly that their projections onto $\M$ on
overlapping 
sets of $\U$ agree giving rise to a globally well defined top 
form  $\tilde{\Omega}$ on 
$\M$. Furthermore we can show that the above expectation values (12) 
and (14) of 
gauge invariant observables $f$ can identically be rewritten as 
corresponding integrals  over the gauge orbit space $\M$ with respect to 
 $\tilde{\Omega}$ 
\beq
\langle f \rangle = 
\frac{ \int_{\M} \tilde{\Omega} \, f} 
{\int_{\M} \tilde{\Omega} }.
\eeq
We note that this last expression shows agreement with the formulation 
proposed by Stora \cite{Stora} upon identification of $\tilde{\Omega}$
with the Ruelle-Sullivan form  \cite{Ruelle}. 
Whereas in \cite{Stora}, however,  this definition of 
expectation values on 
$\M$ appeared as the starting point 
for a global formulation of Yang-Mills theory in the whole space of 
gauge potentials it appears now as our final result. 

We directly aimed 
at the
derivation of  a global path integral formulation in the whole space 
of gauge 
potentials within the stochastic quantization approach; we recall 
that we  first derived a  path integral 
 in terms of the local probability density 
$\mu_{\alpha} \, e^{-S_{\alpha}^{\rm tot}}$ which 
 assured   gaussian 
decrease along the gauge fixing surface
$as \, well \, as$ along the gauge orbits. The inherent 
interrelation 
of the field variables on $\Gamma_{\alpha} \times \G$  subsequently  
led to  simple relations of the local
 densities  in the overlap regions and eventually
to  the global  path integral formulations   (12) and (14) on the whole 
field space, as well as to (17) on the gauge orbit space, 
respectively. 

It is remarkable that   the projections onto the local gauge 
fixing surfaces $\Gamma_{\alpha}$ of in specific the
original Parisi--Wu stochastic process  induce a {\it globally defined
stochastic process on the gauge orbit space} yielding the construction
of the
globally defined  path integral density $\tilde{\Omega}$. 
In our opinion this relationship of the globally defined 
Parisi--Wu Langevin equation on the whole field space to the globally
defined path integral density on the gauge orbit space closes nicely one
of the left open issues of the original paper \cite{Parisi+Wu}.
 
 We are aware that in a mathematically 
 strict sense our results are rather formal due to the infrared 
 and ultraviolet infinities inherent in the path integral; 
 it seems challenging to investigate the 
 applicability of a previously developed stochastic 
 regularization scheme \cite{halpern} within our generalized approach.
 
We thank C. Viallet for valuable discussions. G. Kelnhofer acknowledges 
support by "Fonds zur F\"orderung der wissenschaftlichen 
Forschung in \"Oster\-reich",  project P10509-NAW.

\end{document}